\documentclass{article}[amsmath,amssymb]

\usepackage[margin=1in]{geometry}
\usepackage{lmodern}
\usepackage{float}
\usepackage{graphicx}
\usepackage{dcolumn}
\usepackage{bm}
\usepackage{xcolor}
\usepackage{ulem}
\usepackage{amsmath}
\usepackage{amssymb}
\usepackage{ragged2e}
\usepackage{authblk}
\usepackage[colorlinks=true,citecolor=blue,urlcolor=blue,linkcolor=blue]{hyperref}

\usepackage[%
style=phys,%
doi=false, url=false,%
autocite=plain,%
articletitle=false,biblabel=brackets,%
chaptertitle=false,pageranges=false%
]{biblatex}

 \bibliography{arxiv}

\newcommand{\X}{$X~^1\Sigma^+$}
\newcommand{\I}{$I~(\Omega=1)$}
\newcommand{\Hh}{$H~^3\Delta_1$}
\newcommand{\Q}{$Q~^3\Delta_2$}

\newcommand{\wn}{cm$^{-1}$}
\newcommand{\si}{Hz}
\newcommand{\Msi}{MHz}

\title{Optical cycling on thorium monoxide (ThO) for an improved test of fundamental symmetries}

\author[1]{Alexander Frenett\thanks{frenetta@msu.edu, frenett@frib.msu.edu}}
\author[1]{Dorothy Gan}
\author[1]{Nicholas Emtage}
\author[1]{Monika Fouad}
\author[1]{Sebastian Miki-Silva}
\author[1,2]{Xing Wu}
\affil[1]{Facility for Rare Isotope Beams, Michigan State University, East Lansing, Michigan 48824, USA}
\affil[2]{Department of Physics and Astronomy, Michigan State University, East Lansing, Michigan 48824, USA}

\begin{document}
\maketitle
\justifying

\begin{abstract}\justifying
Optical cycling refers to repeated excitation and spontaneous emission on an electronic transition in an atom or molecule. Optical cycling in molecules can enable a wide range of quantum control and readout techniques, but unfortunately it has only been demonstrated on a small class of alkali-like or alkaline-earth-like molecules. Thorium monoxide (ThO), a molecule used in one of the most precise permanent electron electric dipole moment (eEDM) searches (ACME~\cite{Andreev2018}), does not fall into this category. In this work, we demonstrate the first optical cycling on this non-conventional class over a range of experimental parameter space, including laser intensity, polarization switching rate, and interaction time. We show that both the $J=1,2$ rotational levels of ThO molecule are capable of cycling 11(2) photons on average with a single laser, at $1.9(6)\times10^6$~s$^{-1}$~and $2.3(7)\times10^6$~s$^{-1}$~scattering rate, respectively, before population is lost to other vibronic levels. We outline a scheme to apply this demonstrated optical-cycling in an ACME-style eEDM measurement, improving the detection efficiency by over fourfold compared to non-cycling fluorescence detection. This would lead to over a twofold enhancement in the statistical sensitivity of the eEDM search. This optical-cycling scheme can be further extended to scatter $\sim100$~photons, which would enable a wider range of quantum control and sensing using ThO molecules.
\end{abstract}

\section{Introduction\label{intro}}

Optical cycling is foundational to many breakthroughs in quantum control and sensing of atoms and molecules over the last four decades. The cyclic process of rapid excitation on an electronic transition followed by a spontaneous emission back to the original state allows a single atom or molecule to interact with coherent light many times. With increasing number of cycles, a variety of techniques have been realized with high efficiency, including lower-NA optical detection~\cite{Lasner2018_stat}, transverse cooling~\cite{Joffe1993, Shuman2010, Kozyryev2017_sisyphus,Vazquez-Carson2022,Kogel2025}, deceleration~\cite{Andreev1981,PhillipsMetcalf_slowing, Barry2012_slowing}, magneto-optical trapping~\cite{Raab1987_MOT, Vilas2022}, and internal state manipulation (e.g. internal state cooling~\cite{Manai2012,Wakim2012,Staanum2010,Schneider2010,Kogel2025b,Grasdijk2025}).

The ability to optically cycle is not guaranteed in a atom or molecule. In atoms, arbitrary combinations of ground and excited states do not generally form a cycle: a high density of intermediate electronic states tends to lead to more allowed loss channels. If a transition can be identified where the loss channels are few, additional repumping lasers (``repumpers'') can be added to return population to the main cycle and extend the overall length. Repumping has been applied to extend optical cycling to more complex atoms~\cite{Yoon2010,Inoue2018,Eustice2025}, but remains impractical in many species since angular momentum selection rules are not sufficient to rule out large numbers of decay channels when state density is high.

Molecules can have similar electronic complexity as atoms, but also have rotational and vibrational substructure that increases the state density far over that of even the most complex atoms. It was because of this extreme state density that it took decades until Di Rosa~\cite{DiRosa2004} noted that certain molecules could systematically be expected to be able to optically cycle large numbers of times with a small number of repumpers. In the simplest case, combining alkaline-earth(-like) atoms with halogen(-ic) ligands, one could mimic the simple electronic level structure of alkali atom D$_1$/D$_2$ lines in the highest-occupied- and lowest-unoccupied- molecular orbitals (HOMO/LUMO) to avoid the possibility of electronic branching. Rotational branching in these species was shown to be fully controllable by driving specific rotational subtransitions~\cite{Stuhl2008}. In the simplest of these cases, then, the only loss from the molecular optical cycle occurs from the statistical nature of vibrational decays. Even so, the electronic structure of these molecules ensures a similar bond length in the HOMO and LUMO, which in turn increases diagonal decay (i.e. $v'' = v'$, where $v''$ and $v'$ are the vibrational quantum number of the ground and the excited state of the transition, respectively) probability towards unity. Vibrational loss can hence be limited to a few percent in exceptional species (e.g. SrF~\cite{Shuman2009}, CaF~\cite{Wall2008,Pelegrini2005}, RaF~\cite{Udrescu2024}) , and $<10\%$ in a broad swath of molecules with this form, including polyatomics~\cite{Kozyryev2017_sisyphus,Vilas2022,Augenbraun2020} (up to \textit{at least} six-atoms~\cite{Mitra2020}, and possibly much larger~\cite{Zhu2022,Mitra2022}). Select cases of Group 3 metals bonded to chalcogens (Group 16) have also been shown to have alkali-atom-like structure~\cite{Collopy2015,Yang2016_ScO}. This structural principle has also proven effective for mimicking alkaline-earth-atom-like structures in molecules constructed of Group 3 metals and halogen(-ic) ligands~\cite{Yang2016_AlCl,Hofsss2021,Daniel2021} while maintaining favorable vibrational branching. 

In recent years it has become clear that these are not the only types of molecules that have transitions amenable to optical cycling. Calculations and experimental measurements suggest that a diverse range of species (e.g. CH~\cite{Schnaubelt2021}, AuC~\cite{Stuntz2024}) can contain such transitions, despite electronic structure that decisively does not mimic that of simple atoms. Whether, or how, the experimental realizations of optical cycling in these species differs from the more ``traditional'' demonstrations is not yet well-characterized. More studies of these cases are necessary to understanding possible limitations or extensions of the technology in these apparently coincidental instances.

We here study thorium monoxide (ThO), another such molecule that, despite complex electronic structure, appears to have a possible optical cycling transition. Vibronic branching fraction data taken by Steimle \textit{et al}~\cite{Kokkin2014_ULI} shows that the \X$-$\I~transition decays back to the ground state with $0.91$ probability despite $>10$ intermediate electronic manifolds (most of which differ by only $\Delta\Omega = 0, \pm1$~\cite{Zaitsevskii2025,Paulovi2003,Tecmer2018}). Indeed, the state only has a few decay channels above $1\%$ probability. The transition was also measured to have a $1.84$~D~\cite{Kokkin2014_ULI} reduced transition dipole moment (TDM), similar to other optically-cycled molecules, which allows strong excitation. In some contrast, the $I$ state has a $115(4)$~ns lifetime, about $4\times$ longer than the spontaneous emission lifetimes of the excited states in the conventional optical-cycling cases. 

ThO cycling is of interest both for its unorthodox structure, and for its potential application in a state-of-the-art test of fundamental physics. It is currently used in the ACME experiment~\cite{Andreev2018}, one of the most precise measurements of the electron permanent electric dipole moment (eEDM). ACME provides some of the strongest overall bounds on beyond-Standard-Model (BSM) sources of leptonic time-reversal symmetry ($\mathcal{T}$) violation and limits possible BSM extensions that can help explain the observed matter-antimatter asymmetry in the universe~\cite{Cesarotti2019,Chupp2019}. Optical cycling is not currently employed in the ACME detection scheme, during which only about $10\%$ of the molecules in the beam are detected~\cite{Andreev2018}. Cycling even at the level of $\sim10$ scattered photons could improve detection efficiency by $\gtrsim4$ times, limited by excess noise from stochastically decaying into dark vibronic states~\cite{Lasner2018_stat}, assuming efficiency of other components would remain unchanged. This would lead to over a twofold enhancement in the statistical sensitivity of ACME eEDM search from an upgrade on the detection system alone, or a fourfold reduction in the amount of time needed for the investigation of systematic effects. Cycling around $\sim100$~photons makes feasible sufficient state-preparation and readout efficiency to enable a nearly molecule-shot-noise limited EDM measurement using a cold beam of ThO molecules. Therefore, studying cycling on ThO not only improves our understanding of nontraditional applications of the process, but also opens a route to further enhancement of detection efficiency in one of the cutting-edge tests of fundamental symmetries of the universe.

In this paper, we investigate the proposed cycling transition in ThO and demonstrate cycling of $\sim10$ photons with a single laser. We first outline the relevant characteristics of an optical cycle with simple models. We then experimentally characterize the cycling properties of ThO as a function of laser intensity, laser polarization switching rates, and interaction time. The laser intensity measurements are compared to Monte Carlo and optical Bloch equation (OBE) models to constrain the $X-I$ vibronic branching fraction to $\lesssim2\%$ uncertainty. The results of the interaction time scan, in conjunction with the branching fraction constraint, enable a measurement of cycle length and scattering rate. We find that ThO can cycle $11(2)$ times out of both $J=1$ and $J=2$ levels in the ground state at $\gtrsim2\times10^6$~s$^{-1}$ scattering rate, consistent with expectations from the naive models. Using these results, we outline a scheme to achieve an order of magnitude improvement in the detection efficiency in an ACME-style Ramsey-based measurement. We conclude with an outlook on how optical cycling on ThO can be extended to the $\sim$100 photon level with only a few additional lasers.

\section{ThO Structure and Optical Cycling\label{background}}
Applications of optical cycling typically depend on the number of cycles a molecule undergoes, the rate these cycles occur (``scattering rate''), and how these properties vary with small changes in experimental parameters. In this section, we discuss the first two of these and how they relate to the specific structure of the $X-I$~transition in ThO. We elucidate the third point primarily through comparison between simulations and experimental results (Sec.~\ref{results}).

The specifics of an optical cycle depend on the properties of the levels and transitions involved. In this work, the ground state of the ThO cycle is the electronic ground state, $X ^1\Sigma^+$, well described in a Hund's case (a) basis. It has a $B_X = 0.331967(27)$~\wn~rotational constant~\cite{Wang2011} and a small g-factor of $g_X < 0.001$. The excited state is $I (\Omega = 1)$, a Hund's case (c) electronic state at $T_0 =19538.99$~\wn. It has a similar rotational constant to $X$, $B_I = 0.32869(3)$~\wn~\cite{Kokkin2014_ULI}. The $\Omega$-doubling is characterized by $q = 0.00154$~\wn, and, most relevant to this work, splits the $J' = 1$ opposite parity states by $92$~MHz and the $J' = 2$ states by $276$~MHz. Magnetic moment measurements found that $g_I = 0.526$, suggesting the state is comprised of a combination of Hund's case (a) $^1\Pi$ and $^3\Pi_1$ components~\cite{kokkin2015_character}. As noted in Sec.~\ref{intro}, the $I$ lifetime has been measured as $115(4)$~ns, and the reduced transition dipole moment (TDM) between $X-I$ is 1.84~D~\cite{Kokkin2014_ULI}. More information about both states can be found in literature~\cite{Edvinsson1967,Wentink1972,Kokkin2014_ULI,kokkin2015_character}.

\subsection{Cycle Length}\label{subsec:cyclelength}
The length of an optical cycle is defined as the average number of photons a molecule scatters before it is lost to some state outside the cycle. This concept can be made formal by defining the probability distribution of scattering events for an ensemble of molecules. To begin, it is convenient to define the probability of scattering at least one photon as $p(1)=1$; any molecule that is \textit{never} excited is not relevant to an experiment. Then, if the probability of the molecule returning to the ground state after excitation is $r$, the probability $p(n)$ of cycling at least $n$ times is $p(n) = r^{n-1}$. The probability of scattering \textit{exactly} $n$ photons before being lost, denoted $P$, is correspondingly $P(n) = p(n) - p(n+1) = r^{n-1}(1-r)$. If any number of cycles is allowed, then the maximum average number of cycles is given by 
\begin{equation}
    \langle n \rangle_{max} = \sum_{i=1}^{\infty} ir^{i-1}(1-r)= 1/(1-r).
    \label{perfect_cycling}
\end{equation} As previous authors have noted~\cite{Hofsss2021}, the average is smaller in typical realistic scenarios where there is an experimentally constrained maximum cycle count, $k$ (for instance, imposed by a finite interaction time). In this case, 
\begin{equation}
    \langle n \rangle = kr^{k-1} + \sum_{i=1}^{k-1} ir^{i-1}(1-r) = (1-r^k)/(1-r).
    \label{cycling}
\end{equation}
This converges to the expected infinitely-allowed-cycles limiting case. We note that this convention defines a cycle as a spontaneous emission (alone). This is the same convention used in, e.g., \cite{Kozyryev2016,Hofsss2021}. Defining a cycle as the combination of an excitation and \textit{return to the initial ground state} produces results that differ by 1 in the infinite cycling limit. We opt for the former definition because it is simpler to compare directly to experimental data.

\begin{figure}[!t]
    \centering
    \includegraphics[scale=0.6]{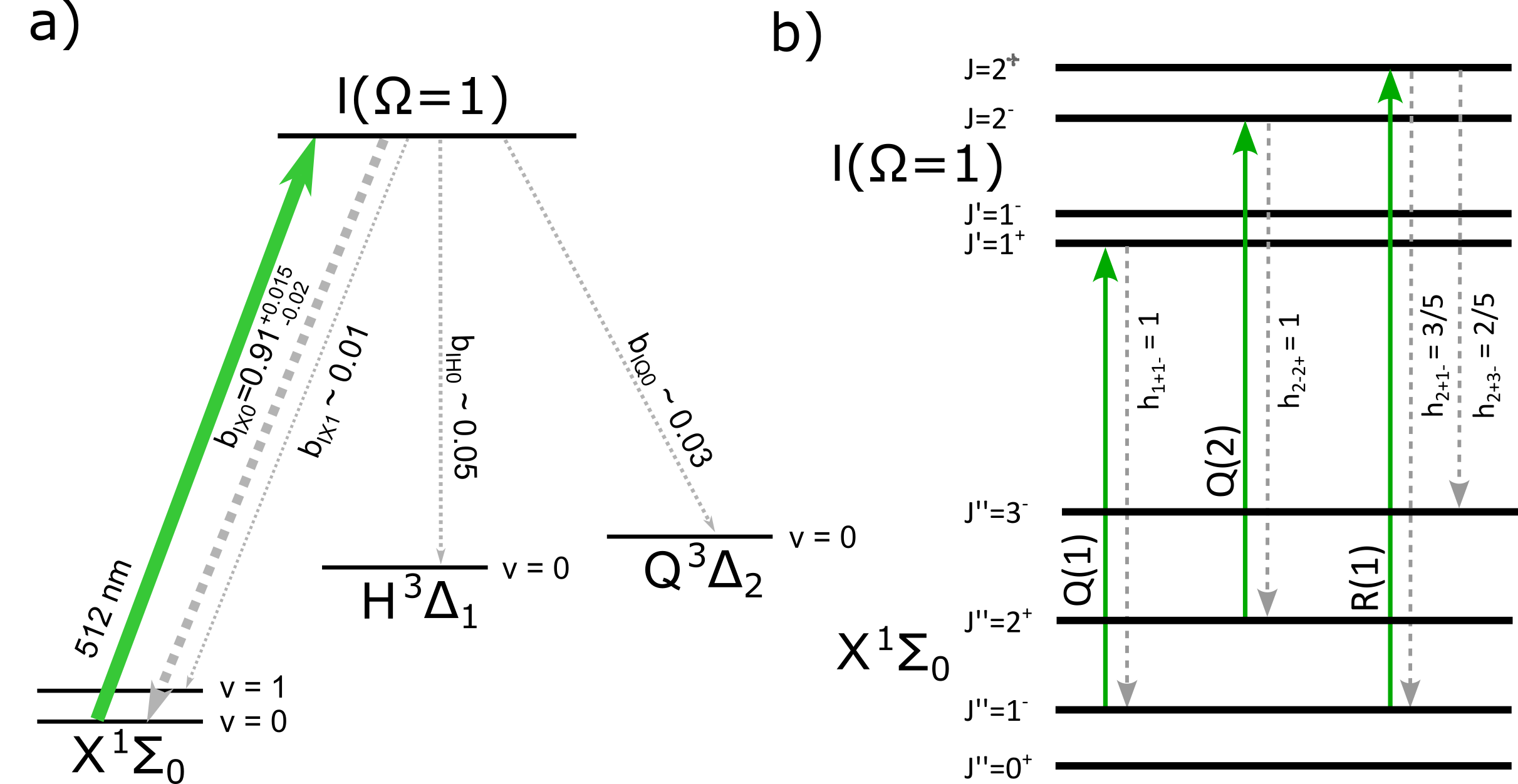} 
    \caption{Level diagrams showing the relevant excitations and branching fractions for both the \textbf{a)} vibronic and \textbf{b)} rotational transitions. The vibronic branching fractions in \textbf{a)} are from~\cite{Kokkin2014_ULI}, and show that $91\%$ of the decay from the $I$ state returns to the vibronic ground state. The uncertainty in $b_{IX0}$ is determined in this work. In \textbf{b)}, the excitations and H\"{o}nl-London (HL) factors $h_{ij}$ for each transition considered in the text are shown. The total branching fraction for a given transition is the product of the vibronic branching fraction and HL factor.}
    \label{fig:branching}
\end{figure}

To characterize an optical cycle length for a transition, the relevant $r$ in the above formulae is given by the product of vibronic and rotational branching fractions. The relevant values for each in ThO are shown in Fig.~\ref{fig:branching}. The vibronic branching fractions (VBFs) are denoted $b_{EG}$, where $E$ is the excited state (presumed to be $v=0$, unless otherwise noted) and $G$ is the ground state, with vibrational state explicit. They are the same for any rotational transition in a vibronic band. As shown in Fig.~\ref{fig:branching}a, the measured~\cite{Kokkin2014_ULI} VBFs from $I$ indicate a $91\%$ probability to decay back to $X, v=0$.

The rotational branching fractions, unsurprisingly, do depend on the rotational transition being driven on a vibronic band. Throughout this paper, we use the common notation $ P(J'')$, $Q(J'')$, and $R(J'')$ for transitions that begin in a ground state with angular momentum $J''$ and change angular momentum by -1, 0, and +1, respectively. In the $X-I$ transition of ThO, $Q$ branches\footnote{Not to be confused with transitions from the \Q~electronic state.} are all \textit{rotationally closed} in the absence of external E fields; the combination of parity and angular momentum dipole selection rules ensure excited population can only decay back to the initial state. In these cases, the vibronic branching fraction is the total branching fraction.

In contrast, driving $R$ or $P$ branches here generally results in less population returning to the initial level, since the excited state $J'=J''\pm1$ can decay back to $J''$ and $J''\pm2$, respectively. The rotational branching back to the original level in these cases is determined by the H\"onl-London (HL) factors, which are tabulated in literature for $^1\Pi/(\Omega = 1) - ^1\Sigma$ transitions~\cite{Hansson2005}, or can be calculated by sums of rotational factors discussed below. The relevant HL factors $h_{ij}$ for transitions studied in this paper are included in Fig.~\ref{fig:branching}b. 

The ability to drive transitions with the same VBF but different HL factor allows us to effectively change cycle length by a factor of 5, from $1/(1-0.91) \approx 11.1$~cycles on $Q(1)$ to $1/(1-0.6\times0.91)\approx2.2$~cycles on $R(1)$. 

\subsection{Scattering rate}\label{subsec:Rsc}
The scattering rate $R_{sc}$ (or cycling rate) is defined as the inverse of the average time to undergo a single spontaneous emission (i.e. to scatter a single photon). It is in general more complicated to model than the cycle length, since it depends on the interplay of several experimental parameters, including the magnitude of the pumping field, transition dipole moment, excited state lifetime, etc. As a starting point, it is well-known that the average scattering rate is strictly bounded to be $R_{sc} \leq n_e/(n_g + n_e) \Gamma$~\cite{FITCH2021157}, where $n_{e}$ ($n_{g}$) is the number of distinguishable excited (ground) states and $\Gamma = 1/(115$~ns) is the spontaneous emission rate. Accordingly, a ThO $Q(J)$ branch scattering rates must be less than $\sim4.3\times10^6$~s$^{-1}$. Such a rate is in theory attainable only at intensities high compared to the saturation intensity $I_{sat} = \pi h c/3\lambda^3 \tau$, where $\lambda$ is the wavelength of the transition and $\tau$ is the excited state lifetime~\cite{FITCH2021157}. For ThO $X-I$, $I_{sat}\approx 1.3$~mW/cm$^2$. 

Even at arbitrarily high intensities, approaching this limiting scattering rate is impossible in ThO with a single polarization of light. The $Q(J)$ lines, like all known molecular optical cycling transitions have $n_e\leq n_g$. Such ``type-II'' systems (at least with integer angular momentum~\cite{berkeland2002}) necessarily have ground states that are dark to an applied laser field of any particular polarization. In this situation, optical pumping quickly accumulates population in the dark state, reducing the scattering rate to near zero after a few excitations. These dark states are typically destabilized either by applying transverse magnetic fields or by rapidly switching the laser field between orthogonal polarizations~\cite{berkeland2002}. The $X$ state in ThO has essentially no magnetic moment, so we must switch the polarization at some rate $R_{ps}$. This requirement introduces another timescale into the scattering rate problem, since all $R_{ps}$ are not equally effective: switch too quickly relative to $1/\Gamma$ and the population does not have enough time to fully populate dark states; too slowly and $R_{sc}$ is strongly dominated by the time sitting in a dark state. To avoid ambiguity in units, we report $R_{ps}$ as 1/$\tau$ in units of \si, where $\tau$ is the switching period in seconds.

To study the effects of $R_{ps}$ on $R_{sc}$ more quantitatively, we use a rate-equation model, similar to as in  Refs~\cite{Wall2008,Hofsss2021,FITCH2021157}. Particulars of the model can be found in Appendix~\ref{App1}, with key details outlined here. Briefly, the model simulates population moving between individual magnetic sublevels in the ground and excited state. The rate of population transfer due to an applied field between a state $i$ and $j$ is given by $R_{ij} = \Omega_{ij}^2/\Gamma$. The Rabi frequency $\Omega_{ij}$ generally depends on the ground and excited state projection quantum numbers for a particular field polarization: $\Omega_{ij}=S_{ij}DE/\hbar$ where $E$ is the electric field strength of the applied field, $D$ is the reduced TDM (1.84 D), and $S_{ij}$ encodes the polarization dependence. In Hund's case (c)~\cite{Wu2020},
\begin{equation}
    S^{\Omega',J',M'}_{\Omega'',J'',M''} = (-1)^{\Delta M+M'-\Omega'}\sqrt{(2J''+1)(2J'+1)}
    \begin{pmatrix}
        J'' & 1 & J'\cr -\Omega'' & -\Delta \Omega & \Omega'\cr
    \end{pmatrix}
    \begin{pmatrix}
        J'' & 1 & J'\cr -M'' & -\Delta M & M'\cr
    \end{pmatrix},
    \label{eq:Sfactor}
\end{equation}
where $\Delta M$ depends on the polarization of the light. Polarization switching is implemented in the model by temporally toggling between $S_{ij}$ calculated with opposite polarizations. Spontaneous emission out of an excited state $j$ occurs at $-\Gamma$, and the spontaneous emission into a ground state $i$ from a specific excited state $j$ occurs at $b_{ji}\Gamma S_{ij}^2/\sum_kS_{kj}^2$ (where $b_{ji}$ is the VBF). The scattering rate can either be extracted from this model as $\Gamma\times(\sum_j e_j)$ as a ensemble-level quantity, or $\Gamma\times(\sum_j e_j)/[(\sum_j e_j)+(\sum_{i} g_i)]$ as a property of an average particle in the cycle, where $e_j~(g_i)$ is the fractional population in the excited (ground) state $j~(i)$. We opt for the latter definition. These definitions converge if $R_{sc}$ is measured on timescales much shorter than $~\langle n\rangle_{max}/\Gamma$, which is often the case with long optical cycles, but not the case here.

We probe a range of $\Gamma,$ $\Omega$, and $R_{ps}$ in Appendix~\ref{App1} to identify combinations of laser intensity and polarization switching rate that generate a high scattering rate. For $Q(1)$, we determine that for $I \gg I_{sat}$, the transition exceeds $2.2\times10^6$~s$^{-1}$~scattering rate when $R_{ps}\gtrsim1$~\Msi. When $R_{ps}<1$~\Msi~, the scattering rate is lower, even more so when $I\lesssim I_{sat}$. We do not explore $R_{ps}>2$~\Msi, but at least until that limit, higher switching rate increases $R_{sc}$ more at high intensities. The fractional increase in $R_{sc}$ generally diminishes over $R_{ps}\approx$1~\Msi.

We find qualitatively similar trends for $Q(2)$, with a slightly faster $R_{sc}$ than $Q(1)$ in typical parameter configurations (exceeding $3.2\times10^6$~s$^{-1}$~at high intensities and $R_{ps}\approx1~$\Msi). This increase aligns with previously reported models~\cite{Hofsss2021} and stems from a smaller proportion of dark states in the higher $J$ level for any given polarization. 

In this section, we have shown that simple models suggest that ThO is able to cycle $\sim11$ photons at $\gtrsim2\times10^6$ s$^{-1}$ scattering rate when polarization switches at around $1$~\Msi. A more complete optical cycle in the future will include repumping lasers that cover more vibronic branching fraction and extend the cycle length. Some repumpers will likely connect more states to the $I$ manifold, decreasing the scattering rate. The estimates derived in this section thus correspond to a lower limit of cycled photons and an upper limit of the scattering rate achievable on the ThO $X-I$ transition overall. 

\section{Apparatus\label{sec:Apparatus}}
\begin{figure}[!t]
    \centering
    \includegraphics[width=\linewidth]{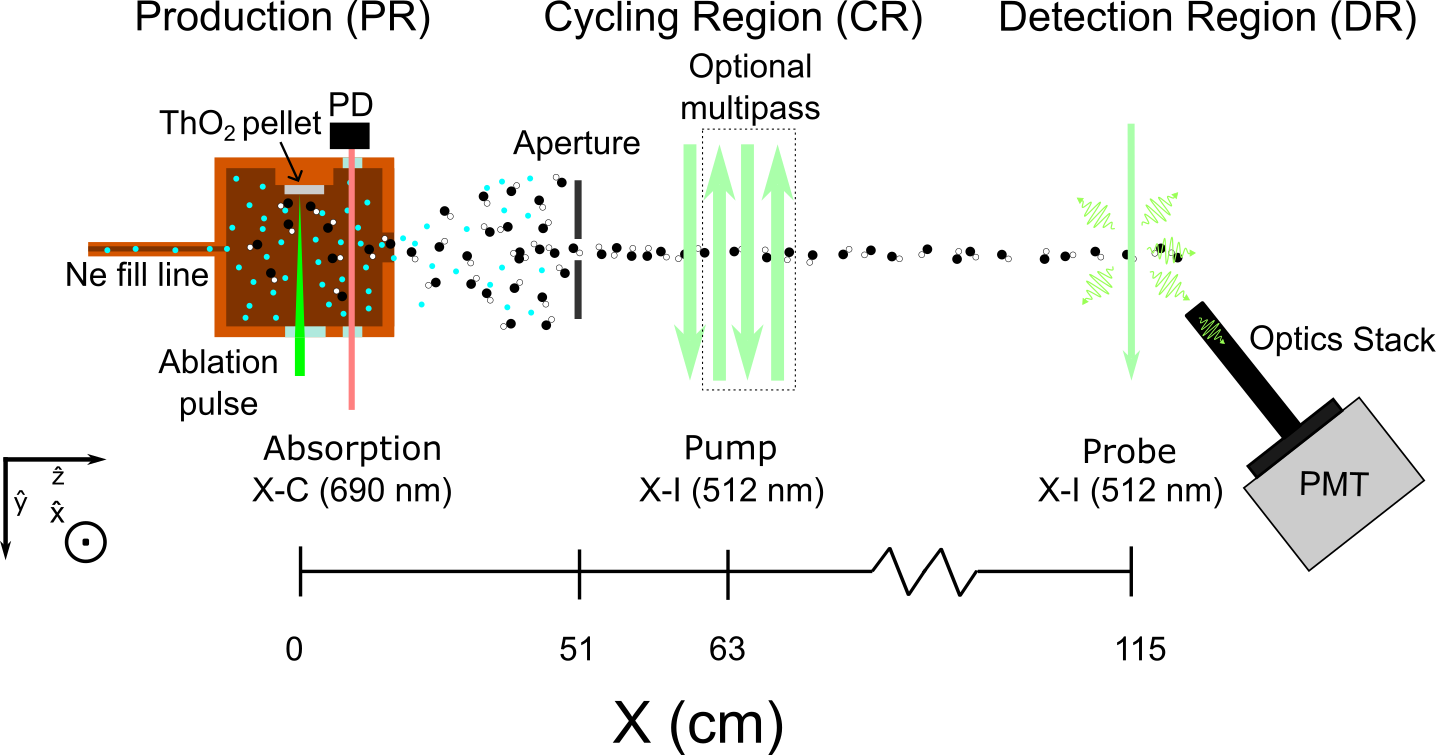}
\caption{Schematic of the experimental setup. A beam of thorium monoxide molecules is first produced in a cryogenic cell via laser ablation of ceramic thorium dioxide. Ambient neon gas thermalizes the molecules and extracts them in a beam. The beam is collimated by an adjustable aperture of $5~\mathrm{mm}\times5~\mathrm{mm}$ or smaller. After a few cm, the molecules are pumped with the cycling light on the $512$~nm $X-I$ transition in the cycling region (CR). The additional multi-pass of pump laser in the dashed box is optional. The molecules are probed downstream, at the detection region (DR). The probe laser is also on the cycling transition. In the DR, molecular fluorescence is collected by an optics stack onto a photomultiplier tube (PMT).  The polarization of both lasers is rapidly switched between orthogonal polarizations to access magnetic dark states. The exact configuration of light in the CR and DR depends on the measurement being taken, as discussed in Sec.~\ref{sec:data}.}
    \label{fig:apparatus}
\end{figure}

To study the cycling characteristics experimentally, we conduct a variety of spectroscopic measurements on a cryogenic buffer gas beam (CBGB) of ThO throughout a 1.15~m vacuum beamline. We describe here the experimental apparatus (Fig.~\ref{fig:apparatus}). The measurements are discussed in Sec.~\ref{sec:data}.

An experimental cycle begins in the ``production region'' (PR) with the creation of a thorium monoxide CBGB~\cite{Hutzler2011,Han2026}. In the CBGB source, a ceramic pellet of thorium dioxide is held in a copper cell at $\sim18$ K. The pellet is ablated with a $\sim30$ mJ, $\sim5$ ns, $1064$ nm Nd:YAG pulse in the presence of $\sim10^{15-16}$/cm$^3$ neon buffer gas. ThO is produced in the ablation plume and subsequently thermalized by the ambient neon gas to near the cell temperature. Though the molecules are refractory, the density of the buffer gas prevents a reasonably-sized fraction of them from hitting the wall~\cite{Hutzler2011}, and they exit the cell entrained in the neon as a $\sim250$ m/s beam in the $\hat{z}$ direction. The velocity is measured via time-of-flight. The ablation occurs at 2 Hz. At this repetition rate, a neon flow rate of 25-30 sccm achieves high extraction efficiency of thorium monoxide from the cell. The production efficiency and extraction timescale are monitored by molecular absorption via a 1~mm laser beam passing through the cell tuned to the $X-C,~Q(1)$ transition. After extraction from the cell, the molecular beam is collimated by in-vacuum razor blades to reduce the transverse Doppler widths and spatial spreads along the $\hat{x}$ and $\hat{y}$ axes (defined in Fig.~\ref{fig:apparatus}). 

The molecules then traverse a high vacuum ($\sim 10^{-7}$ torr) beamline to reach the cycling region (CR), in a 6"x6"x12" Ideal Vacuum cube with 512 nm antireflection-coated windows for optical access. In the pump/probe measurements discussed below, the molecular beam intersects a perpendicular laser beam tuned to $X-I$ transition around 512 nm, $\sim60$~cm downstream from the cell exit aperture. The laser beam is typically elongated along the $\hat{x}$ axis using cylindrical lenses to be taller than the molecular aperture, ensuring the majority of molecules are exposed to near uniform intensity of light. 

After the CR, the molecules continue down the beamline for another 52 cm until they reach the detection region (DR), inside an ISO-100 eight-way cross. The interior of the cross is blackened with a combination of oxidized copper inserts and Acktar black foil. Vacuum is maintained by a 720 L/s turbomolecular pump. In the center of the DR, the molecular beam again intersects by a perpendicular laser light tuned to the $X-I$ transition. This beam is typically $\sim3$~mW and $\sim1$~mm in diameter. The laser-induced fluorescence produced in the DR is collected by an optics stack onto a photomultiplier tube (PMT) at $45^{\circ}$ to the direction of molecule propagation. In the optics stack, fluorescence photons first encounter a two-inch-diameter f=300~mm collection lens positioned at a focal length away from the intersection of the molecule and the probe laser beam. This lens is followed by a $3.3\times$ reducing telescope, a 532 nm notch filter, and a 510/10 nm bandpass filter. This combination spatially and spectrally filters out most background that is not emitted by molecules decaying on the $I\rightsquigarrow X$ line from the vicinity of the laser/molecular beam intersection. ZeMax simulations of the setup indicate a $\sim0.2 (0.1)\%$ collection efficiency. The fluorescence photons finally impinge on a Hamamatsu R7600U-300 PMT, which has a $\approx20\%$ quantum efficiency at 512 nm~\cite{Masuda2023}. The PMT output is amplified by a Stanford Research Systems 560 preamplifier to obtain volt-level voltage signals that are read into a computer by a LabJack T7 data acquisition device. 

Different lasers tuned to the $512$~nm $X-I$ transition are used in the CR and DR. This allows independent frequency control for the laser beam in each region. Up to $100$~mW at the experiment, typically the CR, is supplied by passing $1024$~nm light through a periodically-poled lithium niobate (PPLN) doubling crystal. The infrared input is generated by seeding a ytterbium-doped fiber amplifier (YDFA) with a home-built external-cavity diode laser (ECDL). Lower power light, typically for the DR, is separately generated by a $50$~mW 512~nm ECDL. Polarization switching is added differently for each laser. The high-powered light polarization is switched at up to 2~\Msi~by a QUBIG Pockels cell. The polarization of the ECDL light is switched at up to $\sim1.2$~\Msi~using a pair of acousto-optical modulators (AOM) and polarization optics. In the latter case, the switching frequency is limited by distortions from the finite AOM rise time.

The absorption signal in the buffer gas cell is from $\sim10\,\mu$W 690~nm light generated by a home-built ECDL. In-cell collisions rethermalize the molecules after interacting with this light, so its presence does not affect the internal state distribution of the molecules downstream.

All laser frequencies are referenced to a HighFinesse WS7 wavelength meter, which in turn is stabilized to a 1550 nm HCN-stabilized reference. Typical locking accuracy of 1-3 MHz is achieved using this setup, similar to the natural linewidth of the $X-I$ transition.

\section{Experimental Methods\label{sec:data}}
We conducted three groups of pump/probe-style measurements on the CBGB to characterize the optical cycling. This section reports the experimental methods and parameters. Results of these experiments, comparisons with simulation, and discussions thereof are given in Sec.~\ref{results}.

\subsection{Laser Intensity Scans}
A pump laser in the CR was used to excite molecules on an $X-I$ transition with a range of laser intensities. Molecule population that decayed into other vibronic states was lost from the cycle. We measured the amount of loss from the cycle as a function of the laser intensity. The population remaining in the initial state was probed in the DR by excitation on the $Q(J)$ transition. Since there was $50~$cm (i.e. $\approx2\,$ms time-of-flight) between the pump and probe lasers, spontaneous decay from $I$ had completed, so the observed ground state population was the entire population remaining in the cycle. To characterize the cycle closure, measurements of this type were performed with the pump beam on $Q(1),~Q(2),~\mathrm{and}~R(1)$ lines of the $X-I$ transitions. 

For these data, the molecular beam was collimated to $5$mm$\times5$mm. The pump laser beam was shaped to an elongated ellipse with minor radius $0.8$~mm and major radius $11$~mm. The power was measured before beam enlargement using a Thorlabs S130C power meter, with loss due to surface reflections accounted for later. Both pump and probe lasers were polarization switched at 1.2~\Msi. The probe beam was aligned to the molecular beam and PMT optics by changing the vertical ($\hat{x}$-direction, as defined in Fig.~\ref{fig:apparatus}) angle to maximize DR fluorescence on resonance. The pump beam was subsequently aligned vertically by minimizing DR probe fluorescence on the same transition. 

\subsection{Polarization Switching Rate Scans}
The effect of polarization switching was quantified by similar pump-probe measurements. At fixed intensity, the polarization switching rate was changed over a range from $0$ to $2$~\Msi. At each point in the scan, the input polarization of the pump laser through the Pockels cell was adjusted to maximize the contrast between alternate polarizations to account for shifts in the offset birefringence as the RF frequency changes. The power in the pump beam was measured after this adjustment, and small differences were accounted for with a variable ND filter. The polarization switching rate were scanned for both $Q(1)$ and $Q(2)$ cycling lines. For each, data were taken at three laser intensities to also characterize the combined effect of intensity and switching rate. For consistency, the polarization switching rate of the probe beam was kept constant at 1.2~\Msi. The sizes of the molecular beam, and pump/probe laser beams, were kept the same as in the intensity scan.

\subsection{Interaction Time Scans}
Finally, the cycling efficacy as a function of interaction time over which molecules intersect with pump laser field was studied. This was also done via pump-probe measurement, where the population that remained in the cycle after pumping in the CR was measured in the DR.
For these measurements, the molecular beam was shrunk to $\sim2~\mathrm{mm}\times2\mathrm{mm}$. The pump beam was also shrunk to a 2.4~mm radius circular beam. To better define the interaction distance (and thus interaction time), the pump beam was clipped with razor blades to be 1.2(1)~mm along the molecule propagation ($\hat{z}$) axis. The power of the beam was measured after clipping. The pump beam was multipassed through the CR with two sets of retroreflecting prism pairs. Up to four passes of light, with a few mm between each pass, fit along the CR optical windows. In a multi-pass configuration, the average intensity across the entire set of passes is recorded, accounting for reflective loss from the vacuum windows and mirrors. Operating high above $I_{sat}$, at $\sim38$mW/cm$^2\approx 30I_{sat}$, the effect from the small variation in local intensity over each individual laser passes on the scattering rate should be small. Polarization switching rate was set to $1.2$~\Msi, above which the fractional change in scattering rate was found to be flatted (see Sec.~\ref{subsec:RpsScan}). These scans were taken on both $Q(1)$ and $Q(2)$ lines to compare the relative difference in scattering behavior of these two cycling transitions.

\section{Simulations\label{Simulations}}
To model the first two tests of optical cycling in greater detail than the approximate benchmarks presented in Sec.~\ref{background}, we developed optical-Bloch-equation(OBE)-based simulations of the experiment. The simulations are designed to include all relevant experimental parameters and realistic constraints.

They take two steps. First, a Monte Carlo routine simulates the ballistic propagation of molecules leaving the cell and down the beamline, following the apparatus geometry described in Sec.~\ref{sec:Apparatus}. The spatial and velocity distributions of the molecules that survive through the aperture are then sampled at the point where they intersect with the pump laser in the CR. These samples serve as inputs OBE simulations of molecules interacting with the optical pumping laser field, the second step. 

The OBE simulations begin by using the longitudinal velocity, transverse velocity, and off-axis position of a particle to define a discrete trajectory with time-step of $\sim0.3$ ns. Experimentally measured laser power and beam profiles are then used to calculate the Rabi frequency experienced by the particle at each time-step. The Rabi frequency, positions, and velocities define specific OBEs over each time-step. These OBEs are numerically integrated along the trajectory using the Euler method. The momentum recoils from spontaneous emissions are not considered, since they are too little: ten 512 nm scatters change the average velocity of a ThO molecule by only $\sim10$~mm/s, resulting in a $\lesssim50\,\mu$m change in transverse position upon arrival at the probe laser, and negligible in the longitudinal direction which is dominated by the initial $\sim250$~m/s beam velocity.

The OBEs are constructed over the full basis of magnetic sublevels in the ground and excited states that we probe experimentally. To do so, the Hamiltonian for a given transition, with a single polarization of light, is first constructed in the rotating-wave approximation. The excitations and decays are included with appropriate rotational factors (Eq.~\ref{eq:Sfactor}). The OBEs are then calculated using $\frac{d\rho}{dt}=-\frac{i}{\hbar}[H,\rho]$, and appropriate decay terms (under the convention of Ref.~\cite{FITCH2021157}) are subtracted to account for spontaneous decay. We do not include higher-order effects, e.g. coherence between $M=\pm1~\mathrm{levels~in} ~I, J=1^+$ state. In the experiment, such terms are only generated during the $<5$~ns transient time of Pockels cell switching over the $\sim1~\mu$s switching period, and quickly decay. To integrate polarization switching, we replace the single-polarization OBEs with those corresponding to the orthogonal polarization at the experimental switching rate. Loss from the cycle is registered by introducing a single additional state, and the probability of decaying into it from the excited state is given by $(1-b_{IX0})$, where $b_{IX0}$ is about $0.91$ but can be varied in the simulation to test agreement with measurement. 

The parameters (e.g. collimator size, laser intensity) can be adjusted in the model to compare to the experimental results of the intensity and polarization switching rate scans discussed in Sec.~\ref{results}. In each case, uncertainty in the simulation is obtained by varying model inputs over the range of experimental uncertainties. Since these simulations account for experimental geometry and approximate the CBGB molecular state distribution, they provide a more robust means of quantitative comparison to experimental results than the simple models developed in Sec.~\ref{background}. 

\section{Results and Discussion \label{results}}
\subsection{Laser Intensity Dependence}\label{subsec:IntensityScan}
\begin{figure}[!ht]
    \centering
    \includegraphics[width=\linewidth]{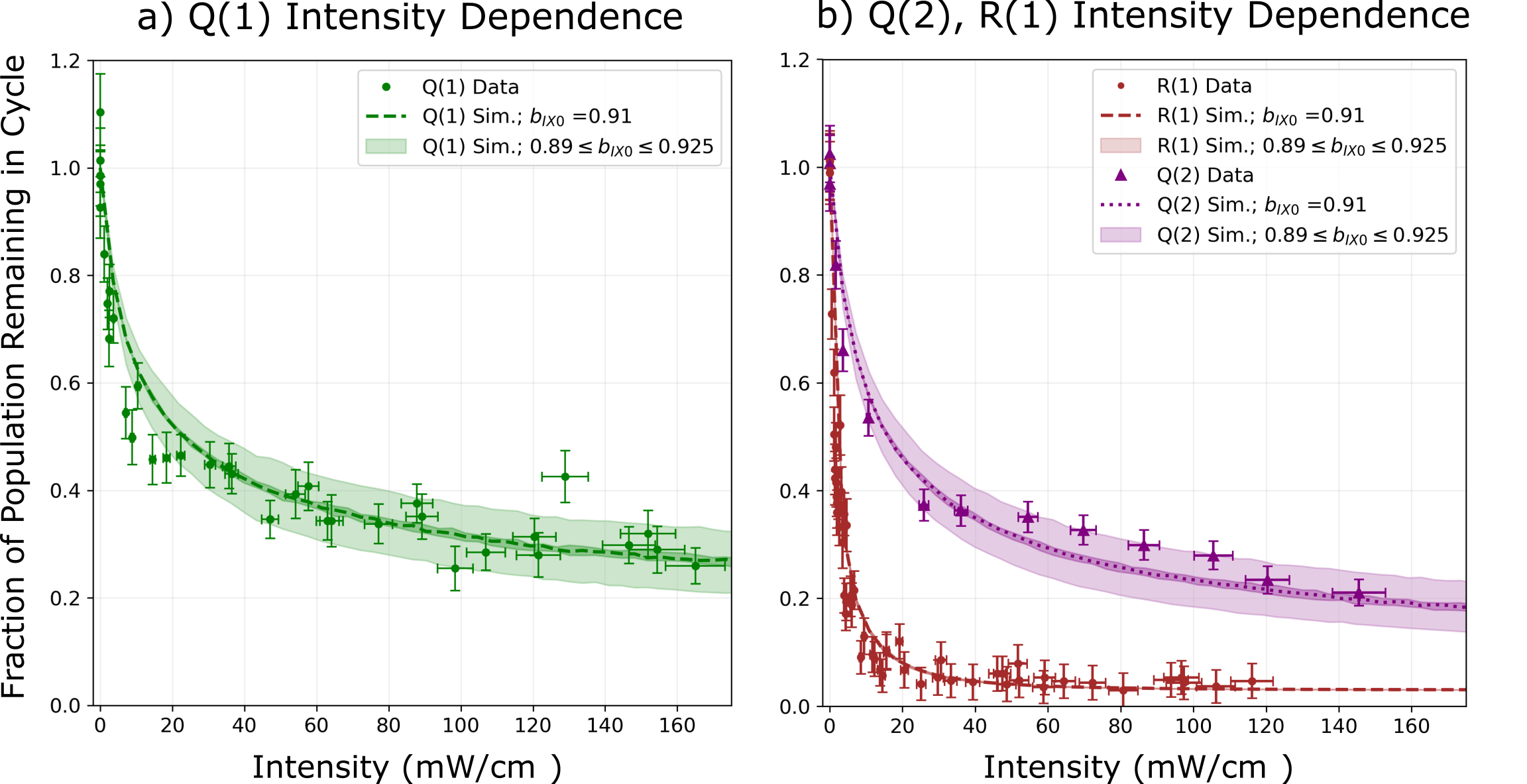}
    \caption{Plots of population depletion out of an optical cycle as a function of laser intensity for three rotational transitions. The laser beam size was kept the same for all datapoints, and the laser polarization was switched at 1.2~\Msi. OBE simulations at the expected experimental parameters are shown in thick dashed and dotted curves. Light (dark) shaded areas indicate ranges of simulation results due to uncertainties in the $b_{IX0}$ VBF (the excited state lifetime). The bounds on the $b_{IX0}$ VBF are determined from the $\chi^2$ value analysis, described in the text. \textbf{a)} On the left, the intensity dependence of the $Q(1)$ (green) cycling transition is shown. Even at high intensity, there remains population in the cycle due to the high branching fraction back to the initial state. \textbf{b)} The same can be seen in $Q(2)$ (purple) on the right. The $Q(2)$ cycle is more efficient than $Q(1),$ owing to the lower fraction of instantaneous dark states. These two closed cycles can be compared to the $R(1)$ plot (brown) on the right. $R(1)$ is not rotationally closed, and has a $40\%$ smaller branching back to the ground state than the closed transitions. The $R(1)$ cycle quickly empties the population, far before the peak intensity is reached. }
    \label{fig:intensity_scan}
\end{figure}

The results of the laser intensity scans, and accompanying simulations, are shown in Fig.~\ref{fig:intensity_scan}. Individual data points are an average of 50-100 experimental shots at the corresponding laser intensity. Sets of data points taken within the same day are normalized, but there is no fit parameter in the simulation models. Error bars represent a sum in quadrature of the uncertainty on the normalization factor, statistical uncertainty, a $\sim5\%$ uncertainty associated with molecular signal fluctuation out of the CBGB, and systematic analysis uncertainty caused by scattered light background. The dotted and dashed curves in Fig.~\ref{fig:intensity_scan} show the results of the OBE simulations with the measured molecular and laser beam parameters. The dark shaded regions indicate uncertainty in simulation from excited state lifetime uncertainty~\cite{Kokkin2014_ULI}. The light shaded region shows the range of simulation results when the VBF is varied by $+0.015/-0.02$, which corresponds to the uncertainty on the $b_{IX0}$ VBF we derive later in this section. There is a 0.03 offset (measured below) included to each simulation curve to account for continuous leakage into $X$ from long lived excited states (e.g. $Q~^3\Delta_2$), which is not accounted for in the OBEs. We note that since the pump laser beam is extended in the $\hat{x}$ direction, fluorescence due to off-axis molecules that are non-resonant with the pump beam should be small in this configuration.

The impact of branching fraction is clear by comparing the intensity scans of the (closed) $Q$ lines to that of the lossy $R(1)$. In the $R(1)$ data, the population is almost entirely pumped out of the $X, J=1$ state when the pump beam $I\gtrsim I_{sat}$. In contrast, the $Q$ lines do not fully empty the ground state in this measurement; even at over $I=100I_{sat}$, there is still $>20\%$ of the original population remaining. Since $Q(1)$ and $R(1)$ drive out of the same state, this difference cannot be explained by different ground state multiplicities. Some part of the difference is made up by excited state multiplicity. Since in the $R(1)$ transition there are fewer ground states than excited states, the scattering rate is allowed to be up to $2.5\times$ faster than that in $Q(1)$, though this is only possible near the upper intensity limit on the plot. Since the $Q(1)$ curve does not match the $R(1)$ curve stretched along the x-axis by a factor of 2.5, the majority of the difference is caused by the different branching fraction back to the ground state. On the $R(1)$ curve we can also see a clear offset ($\sim0.03$) from 0 in the tail. Since the ``cycled'' molecules are all lost from the process, this background is almost entirely due to decays into $X$ from long-lived excited states~\cite{Ang2022,Wu2020}, which take place during the time-of-flight between the pump and the probe stages.

It is difficult to extract an average experimental scattering rate in these data, since the interaction time is not well-defined through the gaussian profile of the pump beam. Qualitatively, we note that since the scattering rate must be proportional to the population lost from the cycle, the cycling data are in agreement with the predictions of the rate-equation model that the scattering rate should not strongly increase with intensity past $I=100 I_{sat}$. They also demonstrate agreement with the prediction that the $Q(2)$ scattering rate should be slightly faster than $Q(1)$.

There is, however, sufficient information to derive an experimental bound on the VBF of the $X-I$ transition using the data and simulation results. We compute the combined $\chi^2$ values from comparing both $Q(1)$ and $Q(2)$ data to their respective simulation results over a range of $b_{IX0}$ to assess the best match. The minimum reduced $\chi^2$ between simulation and data, $\chi_0^2$, is reached at $r=0.91$, consistent with the VBF measured in literature~\cite{Kokkin2014_ULI}. Our $1\sigma$ confidence interval on $b_{IX0}$ extends to the values where $\chi^2=2\times\chi_0^2$ (roughly equivalent to having inflated the error bars to make $\chi_0^2=1$), giving an $b_{IX0}=0.91_{-0.02}^{+0.015}$ result and bound. We note that this VBF measurement is somewhat more stringent than the value previously reported, since it is sensitive to the absolute branching probability from $I$ to $X$ out of \textit{all possible decay channels} from I state, not just the fraction of such decays that occur over a finite energy range (i.e. limited by the finite bandwidth of the detector in the dispersed fluorescence measurement~\cite{Kokkin2014_ULI}). In particular, this measurement constrains the sum of all previously unmeasured decays, including $I-A-X$ and $I-B-X$ channels, to be $\lesssim2\%$ probability. The light shaded regions of Fig.~\ref{fig:intensity_scan}~show the range of simulation result over the range of $0.89\leq b_{IX0}\leq 0.925$. Both $Q(1)$ and $Q(2)$ transitions are clearly consistent with this range. The same shaded region is plotted for $R(1)$, but essentially overlaps the main simulation curve. The rotational branching in the latter case dominates such that the small change of VBF over the simulated range essentially does not change the cycling behavior.

\subsection{Polarization Switching Dependence}\label{subsec:RpsScan}
\begin{figure}[t]
    \centering
    \includegraphics[width=\linewidth]{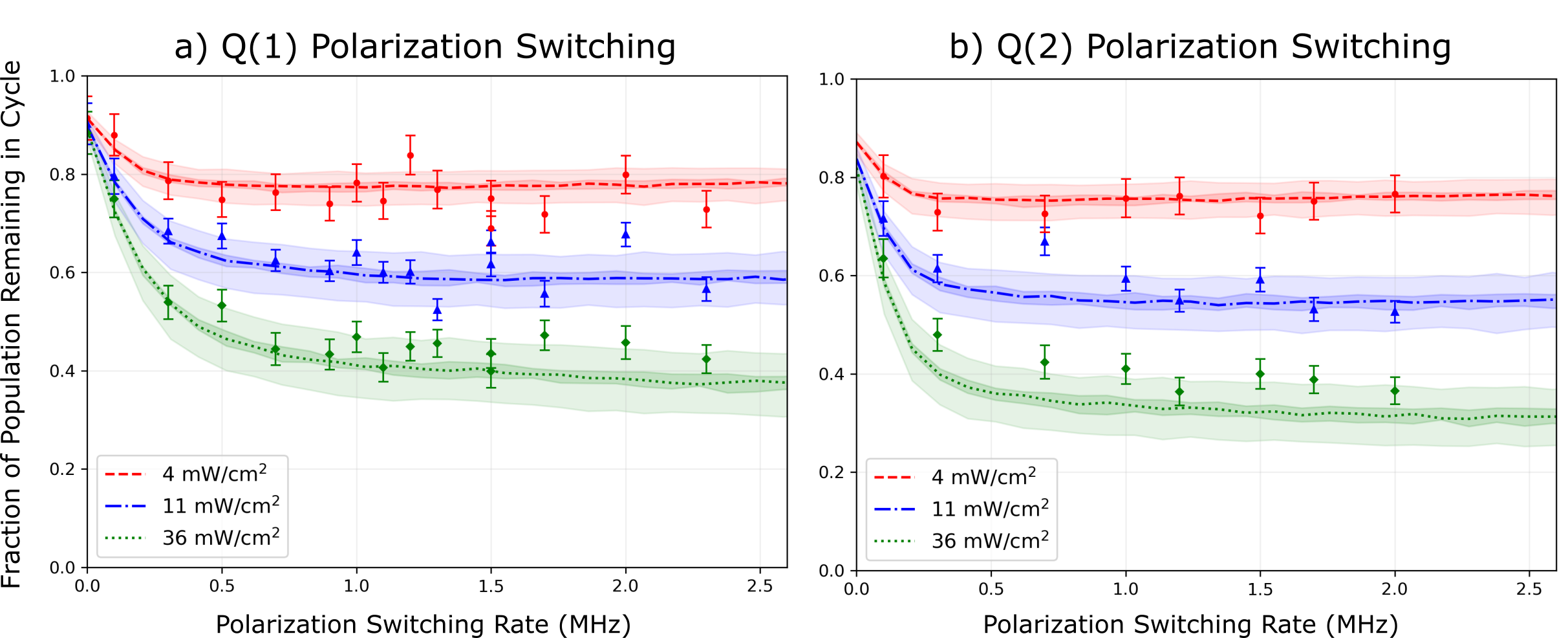}
    \caption{Plots showing the effect of polarization switching on depletion from both \textbf{a)} $Q(1)$ and \textbf{b)} $Q(2)$ optical cycling transitions. The scans over switching frequency were taken at three different intensities on both transitions. Simulation results for each configuration are shown in thick lines. Light (dark) shaded areas indicate ranges of simulation results due to uncertainties in the $b_{IX0}$ VBF (the laser intensities and the excited state lifetime). The uncertainty on $b_{IX0}$ is discussed in Sec.~\ref{subsec:IntensityScan}. Discussion of the polarization switching effects is given in the text.}
    \label{fig:freq_scans}
\end{figure}

The results of the polarization switching rate, $R_{ps}$, scans are shown in  Fig.~\ref{fig:freq_scans}, with accompanying simulations. Each data point consists of sets of 50-100 experimental shots averaged together. For each plot, the entire set of data (over the three intensities) is vertically rescaled by a single parameter to compare to simulation of the same intensities. Error bars on the data points are sums in quadrature of the uncertainty in the rescale parameter, molecular signal noise due to ablation, and noise caused by scattered background light. The standard error in the statistical noise of these datasets was $\sim10\times$ smaller than the smallest of the above mentioned uncertainties. The simulation results at the corresponding laser intensities and switching rates are shown as thick dotted and dashed lines. The dark shaded regions correspond to simulations with laser intensities and excited state lifetime varied by $\pm1\sigma$ from the mean values. The light shaded region corresponds to simulations run over the range of $0.89\leq b_{IX0}\leq0.925$, the limits derived from the laser intensity dependence measurements (Sec.~\ref{subsec:IntensityScan}).

Most clearly, these data show that polarization switching has a qualitatively similar effect on the $Q(1)$ and $Q(2)$ transitions. Even at low $R_{ps}$, the population in all curves is more efficiently depleted compared to no polarization switching, indicating a higher average $R_{sc}$. The increase in depletion, hence in $R_{sc}$, plateaus for the lower intensities by $R_{ps}=0.5$~\Msi~switching rate. At higher laser intensity, the scattering rate flattens around $R_{ps}=1$~\Msi, although continues to increases sublinearly past this point. 

Two differences between the cycling lines can also be seen by comparing the plots. Both are results of the higher fraction of bright states when cycling on the $Q(2)$ line. First, we again see a more efficient cycling process occur in $Q(2)$ than $Q(1)$. It is easiest to see in the saturated region $R_{ps}\gtrsim1$~\Msi, where the curves have flattened: there is less remaining population on the $Q(2)$ plot for every intensity, indicating a reliably faster scattering rate on $Q(2)$ in all comparisons. This is the same effect as observed in Sec.~\ref{subsec:IntensityScan}, and is expected from general arguments in Sec.~\ref{background}. To reiterate, the population freed from a dark state upon a switch in polarization on the $Q(2)$ transition will scatter more times before becoming dark again than when cycling on $Q(1)$, leading to a higher average scattering rate. Both the $Q(2)$ high-intensity data and model indicate more depletion than on $Q(1)$, though the measurements and simulation are in slight tension. This tension indicates that the high scattering rate is somewhat tempered by experimental systematics, a common observation in molecular optical cycling demonstrations~\cite{Williams2017,Collopy2018,Mitra2020,Vazquez-Carson2022}. 

The second difference is seen in the trends of the low-$R_{ps}$ depletion; specifically, in comparison of the 36 mW/cm$^{2}$ intensity data on each transition. Below $\sim0.5$~\Msi, the slope of the $Q(2)$ curve (and data) is significantly steeper than $Q(1)$; not only is the scattering rate faster, but it is increasing more quickly as well. This observation can also be explained by the different multiplicities: since molecules on $Q(2)$ scatter more times before entering a dark state, the relative increase in scattering rate due to an increased $R_{ps}$ should be \textit{larger} when intensity is high and $R_{ps}$ is low, since in these conditions the overall $R_{sc}$ is strongly dominated by dark-state occupation time. This difference could contribute to systematic errors if an experiment requires cycling on both transitions in this parameter regime; small experimental imperfections in $R_{ps}$ would become exaggerated. Operating with $R_{ps}\gtrsim1$~\Msi~reduces sensitivity to this systematic, and is generally both possible and desirable for applications.

In this analysis, we thus find general similarity between the effect of polarization switching on two cycling lines. Where there is difference, we note that the relative impact on experiment can be minimized by operating at sufficiently high $R_{ps}$. Nevertheless, the difference in scattering rate on $Q(1)$ and $Q(2)$ is resolvable in these measurements, emphasizing the necessity of accounting for it when use of multiple cycling transitions is required.

\subsection{Interaction Time Dependence}
\begin{figure}[!t]
    \centering
    \includegraphics[width=0.7\linewidth]{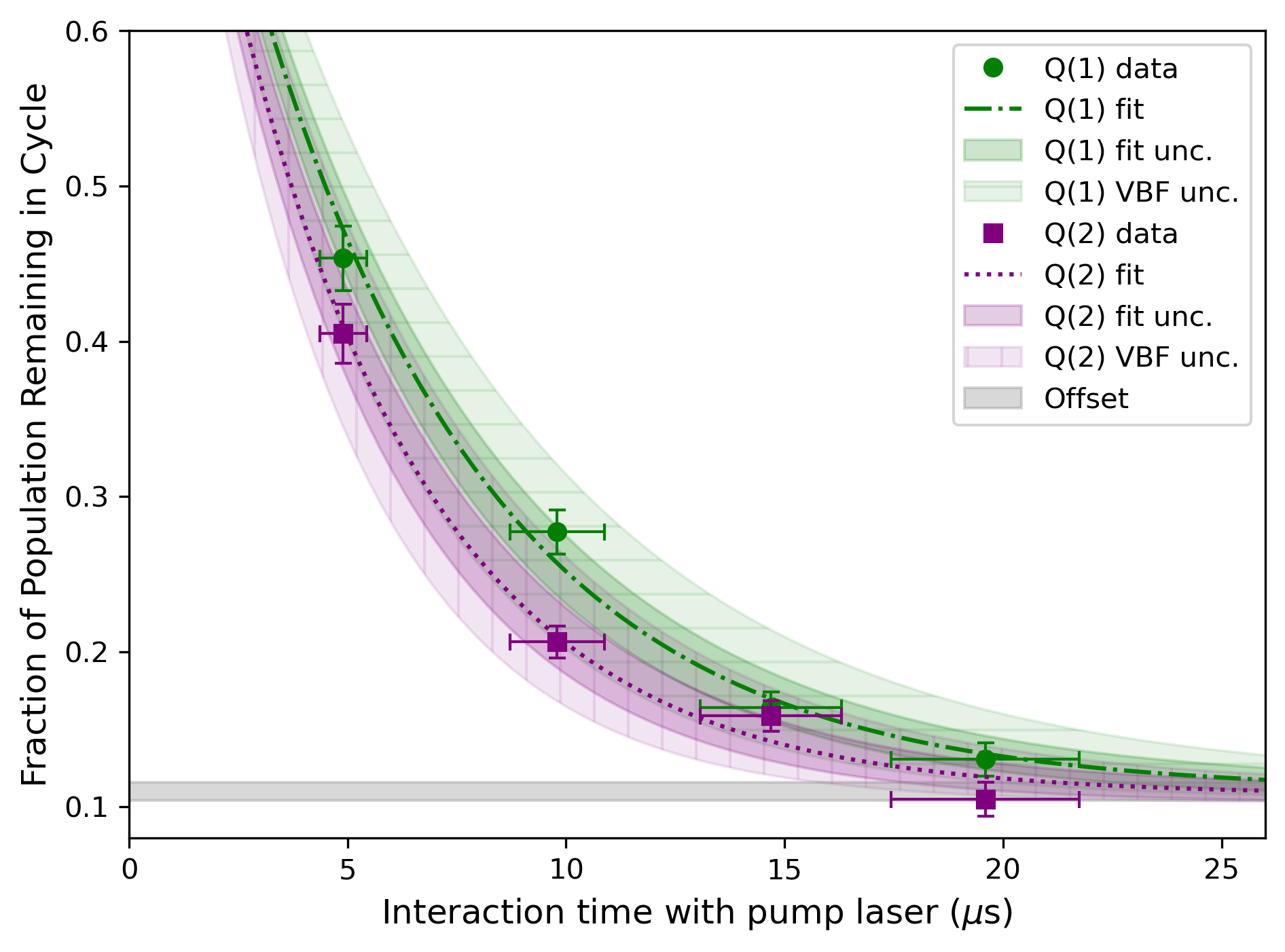}
    \caption{Comparison of $Q(1)$ and $Q(2)$ cycling over a range of interaction times. The remaining population in the ground state decreases with time as $r^{R_{sc}t}$ where $r$ is the branching fraction and $R_{sc}$ is the scattering rate. The interaction time is varied by increasing the number of laser passes that intersect the molecules. The $Q(1)$ data (green circles) fit to a $1.9(6)\times10^6$~s$^{-1}$ scattering rate, and the $Q(2)$ (purple squares) to $2.3(7)\times10^6$~s$^{-1}$. The fits are shown in the dashed and dotted lines, respectively, and also include an offset due to scatter from molecules off the center beam axis. The fit value for the offset is equal for both curves, and represented by the solid gray line. The dark shaded regions show the fit uncertainties of the central value. The light shaded regions also cover the range of uncertainty in VBF, and are therefore correlated between $Q(1)$ and $Q(2)$ curves. The fits suggest that the longest interaction time points represent scattering $11(2)$ photons/molecule on average.}
    \label{fig:int_time}
\end{figure}

The results of the interaction time scan can be seen in Fig.~\ref{fig:int_time}. Each data point is the ratio between the averaged signal from 100 ablation shots with the pump beam present and another averaged signal from 100 ablation shots with the pump beam blocked, taken consecutively. This ratio represents the fraction of the detectable population that remains in the X state after pumping for a given interaction time. Vertical error bars are a sum in quadrature of shot noise, molecular production noise, and scattered light noise. The interaction time axis was obtained from multiplying the number of passes of the pump beam with the clipped pump beam width and dividing by the molecular beam mean velocity. The horizontal error bars represent a sum in quadrature of the uncertainties in the clipped laser beam width and in the molecular beam velocity. In this section, the data are not compared to simulations, but are fitted to extract a cycle depth and scattering rate.

The interaction time scan was set up to simplify the data interpretation and to enable a fit to the simple models developed in Sec.~\ref{background}. Since interaction time is bounded in these measurements by sharply clipping the pump laser light, the maximum number of photons that can be scattered is limited to some value $k$. This simplification enables the average number of photons to be extracted using Eq.~\ref{cycling}. In the experiment, we detect the remaining population in the cycle, which in the model is given by $r^k$, where $r$ is the probability of returning to the ground state after excitation. We parameterize this equation as $r^{qx}$, where $x$ is the number of passes of the pump beam and $q$ is the maximum scattered photons per pass. Here, $r=b_{IX0}$, since the transitions are rotationally closed. To also extract scattering rate, the formula can be again rewritten as $r^{R_{sc}t}$ where $t$ is the interaction time given by clipped laser beam width divided by molecular beam velocity.

We expect a (different, and potentially larger than in Fig.~\ref{fig:intensity_scan}) background in the signal of this depletion scan due to imperfect alignment between the smaller collimating aperture and the pump laser beam of a reduced size. This experimental imperfection allows some molecules to be detected in the DR without having interacted with the pump laser. Since the multipasses are aligned to the same axis as the initial pump beam, these off-axis molecules will not be pumped by any number of followup passes of the pump light. Thus, the the magnitude of the background signal can depend, to the first order\footnote{The different scattering rate between the two transitions will only differentiate the background at higher order, since even when highly saturated, the rate only differs by $\sim20\%.$}, only on the probe laser intensity. As the probe intensity was constant between the $Q(1)$ and $Q(2)$ measurements, we can account for this effect by adding a single offset parameter to the fit.

To then derive scattering rate and cycle length from this model, we vary the value of $r$ used in the fit, in steps of $0.05$ over the range of $0.89\leq r\leq0.925$, and fit the other parameters. We average the fit results over the range of allowed $r$, and combine the uncertainty of the center $q$ value in quadrature with the spread.

From the fits, we find that $q=9(2)$~scatters/pass for $Q(1)$, and $11(3)$~scatters/pass for $Q(2)$. This means that the four-pass data points scatter up to $N_{Q(1)}=36(8)$ and $N_{Q(2)}=44(12)$ photons, respectively. Using Eq.~\ref{cycling}, these results indicate that the average number of photons scattered per molecule in these conditions is $11(2)$ for both $Q(1)$ and $Q(2)$. These data are consistent with the theoretical limit of the maximum cycle length given by Eq.~\ref{perfect_cycling}, $11(2)$ photons, using the measured VBF and its uncertainty range.

Converting $q$ to the scattering rate gives $R_{sc}=1.9(6)\times10^6$~s$^{-1}$ on $Q(1)$, and $2.3(7)\times10^6$~s$^{-1}$ on $Q(2)$. These are consistent with the predictions of the simple rate equation model: $2.2\times10^6$~s$^{-1}$ and $3.0\times10^6$~s$^{-1}$, respectively, as described in Sec.~\ref{subsec:Rsc}. The significant uncertainty is caused by the finite longitudinal and transverse velocity spread of the molecular beam and the uncertainty in the VBF measurement in Sec.~\ref{subsec:IntensityScan}. Smaller uncertainty in the cycle depth, scattering rate, and VBF could be achieved in the future by increasing the optical access in the CR.

The dark shaded regions in Fig.~\ref{fig:int_time} show the uncertainty on the center fit value. The dark regions of the $Q(1)$ and $Q(2)$ curves are uncorrelated. The light shaded regions show the range of fits assuming different $r$ values, and are completely correlated between $Q(1)$ and $Q(2)$. For example, if $r=0.89$, the ``real'' curves will be towards the lower end of the light shaded region, and vice versa if $r=0.925$. Thus, despite the large uncertainty on the center value of $R_{sc}$, the data do unambiguously show that scattering for $Q(2)$ is faster than $Q(1)$, consistent with our observations from the other datasets. This measurement also shows, however, that at modestly high intensity and reasonable switching rate, the same average cycle length can be reached on both rotational cycling lines in $\sim20\mu$s. Simulations suggest the scattering rate could be enhanced further by increasing intensity and polarization switching rate, if necessary, though $20\,\mu$s interaction time is not experimentally difficult to achieve in most contexts.

Through analysis of the interaction time scans and the constraints on $b_{IX0}$ derived in earlier measurements, we confirm the ability to cycle $\gtrsim10$ photons on $X-I$ at about $2\times10^6$~s$^{-1}$ scattering rate on two different rotational transitions. 

\section{Optical-cycling-based Detection Scheme for EDM Experiments\label{cyclingdetection}}

\begin{figure}[!ht]
    \centering
    \includegraphics[width=0.9\linewidth]{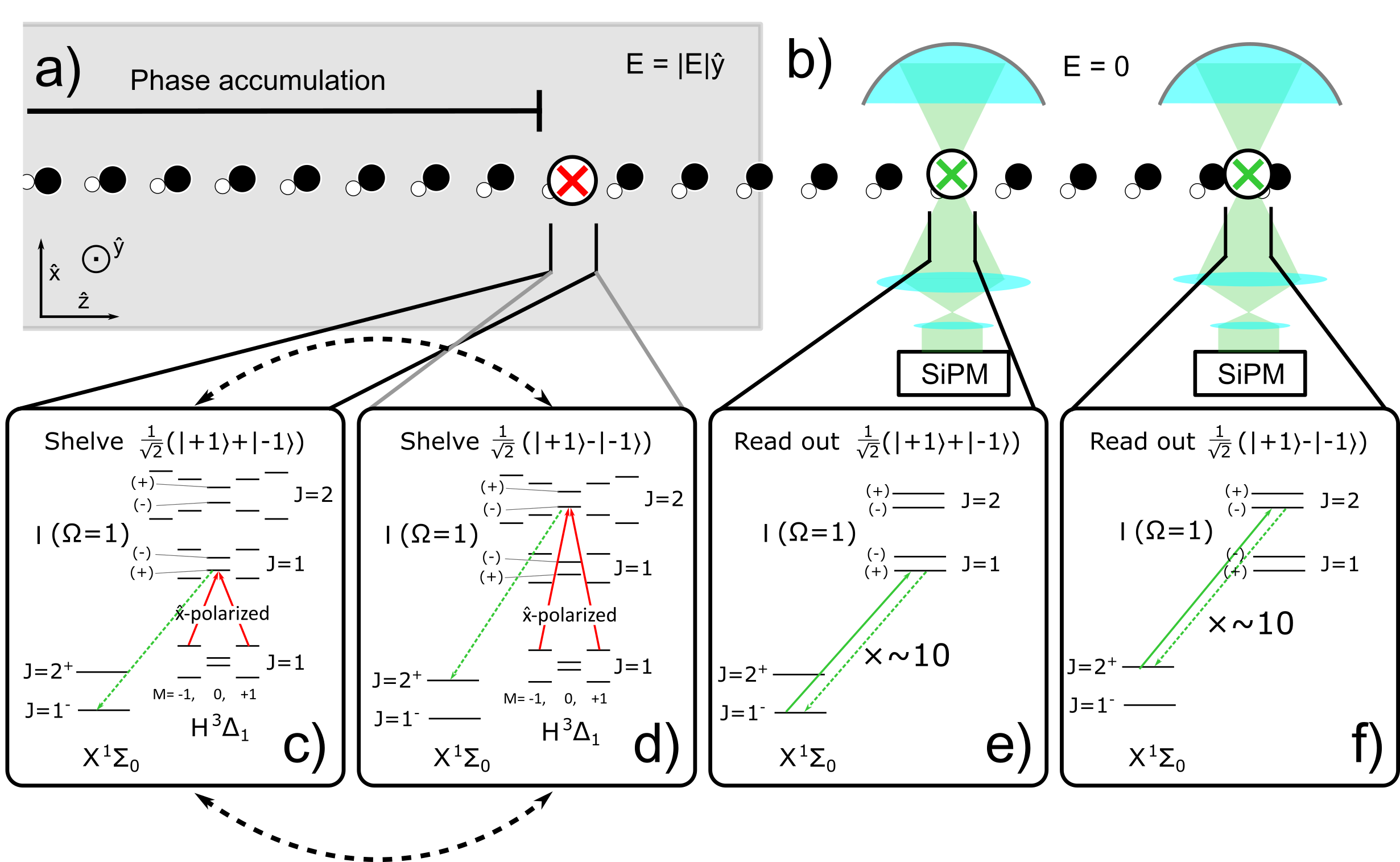}
    \caption{Proposed scheme for optical-cycling detection in an ACME-style electron EDM measurement~\cite{Andreev2018}. \textbf{a)} EDM phase precession region. \textbf{b)} Cycling detection region. \textbf{c)} and \textbf{d)} Level diagrams showing the scheme to shelve $\frac{1}{\sqrt{2}}(|+1\rangle+|-1\rangle)$ and $\frac{1}{\sqrt{2}}(|+1\rangle-|-1\rangle)$ quadratures in the science state into $X, J=1^-$ and $X, J=2^+$ levels, respectively. The bi-directional dashed-arrows represent the application of fast frequency switching to make projection measurement on both quadratures, consecutively. \textbf{e)} and \textbf{f)} Level diagrams showing the scheme to read out populations shelved into $X, J=1^-$ and $X, J=2^+$ levels via optical-cycling transitions, respectively.}
    \label{fig:cycling_detection_scheme}
\end{figure}

One immediate application of the optical-cycling on ThO is to improve the detection efficiency in one of the most advanced eEDM searches. The ACME collaboration~\cite{Andreev2018} uses a cold beam of ThO molecules and is currently working towards a third generation electron EDM measurement with a series of upgrades~\cite{Panda2019,Masuda2021,Wu2022,Ang2022,Masuda2023,Han2026}, aiming to improve upon the current best limit reported by Ref.~\cite{Roussy2023}. In the ACME measurement, ThO molecules are first prepared into the EDM-sensitive science state, a superposition of $M=\pm1$ levels in the $H, J=1$ state. The molecules then undergo a spin precession in a region of well-controlled applied electric and magnetic fields, during which time they accumulate a spin precession phase (Fig.~\ref{fig:cycling_detection_scheme}a). At the end of the precession region, ACME reads out the phase by optically pumping the $H, J=1, M=\pm1$ superposition state to $I, J=1^+, M=0$ state using a laser propagating along the $\hat{y}$-direction (the direction of externally applied electric field, as defined in Fig~\ref{fig:cycling_detection_scheme}a, and collect the fluorescence photons from $I\rightsquigarrow X$ spontaneous decay. In this way, by alternating the polarization of the readout laser between two orthogonal linear polarizations in the $xz$-plane (as defined in Fig.~\ref{fig:cycling_detection_scheme}a, e.g. along the $\hat{x}$ and $\hat{z}$ polarizations, the population in the $\frac{1}{\sqrt{2}}(|+1\rangle+|-1\rangle)$ and $\frac{1}{\sqrt{2}}(|+1\rangle-|-1\rangle)$ quadratures~\cite{Andreev2018}, respectively, are read out via the time-binned fluorescence signal. The efficiency of this detection method is $\approx10\%$, given by the product of the VBF of $X-I$ transition (verified to be $b_{IX0}=0.91^{+0.015}_{-0.02}$ in this work), the photon collection efficiency (about $0.25$), and the quantum efficiency of the SiPM detectors (about $0.45$)~\cite{Masuda2023,Hiramoto2023}.

It is now conceivable to make a ten-fold improvement in the detection efficiency of ThO using the optical-cycling we demonstrated in this work. Fig.~\ref{fig:cycling_detection_scheme} illustrates the plan. Instead of projecting the $M=\pm1$ superpositions via the same excited state but orthogonal linear polarizations of the readout laser as in ACME, we can alternate the readout laser frequencies between going to two different excited states, i.e. $I, J=1^+, M=0$ and $I, J=2^-, M=0$, but with the same linear polarization (Fig.~\ref{fig:cycling_detection_scheme} c,d). This will selectively read out population in the $\frac{1}{\sqrt{2}}(|+1\rangle+|-1\rangle)$ and $\frac{1}{\sqrt{2}}(|+1\rangle-|-1\rangle)$ quadratures in the science state and shelve them into $X, J=1^-$ and $X, J=2^+$, respectively. (The underlying principle for phase projection, coherent population trapping, is the same as used in ACME readout~\cite{Andreev2018}.) After molecules leave the phase accumulation region, they will traverse two consecutive photon collection regions, where we can read out the population in $X, J=1$ and $J=2$ via $Q(1)$ and $Q(2)$ cycling transitions, respectively (Fig.~\ref{fig:cycling_detection_scheme} e,f). This is very similar to the cycling detection scheme used in the YbF EDM experiment~\cite{Ho2020}. 

This plan also comes with built-in features to reject systematic effects in the phase measurement. By changing the linear polarization of the readout laser to its orthogonal direction (e.g. from $\hat{x}$ to $\hat{z}$ directions in Fig.~\ref{fig:cycling_detection_scheme} a), we can swap the two quadratures of the superposition state that get shelved into the $X, J=1,2$ states. Thus, this can act as an experimental switch. The order of the $Q(1)$ and $Q(2)$ cycling readouts can also be swapped in Fig.~\ref{fig:cycling_detection_scheme} e,f, as another experimental switch. 

With the $X-I$ transition laser alone, we can increase the detectable fluorescence photons per ThO molecule from $0.91$ to $\gtrsim10$. This would increase the detection efficiency by around 4 times assuming all other conditions remain the same\footnote{The improvement is less than 10 because there is broad stochastic noise on the number of cycles per molecule~\cite{Lasner2018_stat}}. This would lead to a twofold enhancement in the statistical sensitivity of ACME eEDM search, or a fourfold reduction in the time needed for the investigation of systematic effects in the measurement. This improvement alone is equivalent to hypothetically increasing the collection optics coverage in a noncycling measurement to almost $4\pi$ sr solid angle~\cite{Lasner2018_stat}. With four more repumpers that could help recover population from dominant vibronic decay channels (Fig.~\ref{fig:branching} a), we could possibly cycle $\sim100$ photons per molecule. This provides a pathway towards molecular-shot-noise limited detection in electron EDM searches with ThO molecules.

\section{Conclusion}
In this paper, we achieve cycling of $\gtrsim10$~photons per molecule on ThO using the \X$-$\I~transition without any vibronic repumpers. We begin by setting up simple statistical and rate-equation models to estimate the cycling characteristics. We then conduct three pump-probe experiments on a CBGB of ThO to characterize the cycling scheme. First, the dependence of cycling on pump laser intensity is investigated over a range of rotationally open and closed transitions. The dependence of two cycling transitions on polarization-switching frequency is also measured. These initial measurements identify experimental parameters that ensure near-maximal scattering rate. Optical-Bloch-equation-based simulations are developed to model the measurements with realistic experimental parameters and constraints. The simulations are compared with the measurement results to constrain the $X-I$ vibronic branching fraction to be $91^{+1.5}_{-2}\%$. We then assess the effect of varying the interaction time that molecules traverse the pump laser at a fixed intensity and polarization switching rate. This measurement allows us to extract the average number of photons scattered and average scattering rate on both $Q(1)$ and $Q(2)$ cycling transitions. Both transitions are found to cycle $11(2)$ photons, in agreement with the estimated photons for an infinitely long interaction time assuming the measured value of VBF. Each transition scatters at about $2\times10^6$~s$^{-1}$ rate. These results are consistent with simple statistical and rate-equation models of the cycling process. Finally, we outline an experiment scheme that utilizes the optical cycling to improve detection efficiency by over $4\times$ for a Ramsey-type measurement, such as the ACME eEDM search. This proposed scheme requires only the results demonstrated from this work, which emphasizes the utility of even a small degree of optical cycling. Moreover, the demonstrated optical cycling could be a crucial step towards optically detecting the rare isotopologues, $^{227,229}$ThO,  where only trace amounts of the radioactive isotopes are available. Such applications include a promising nuclear Schiff moment search using $^{227,229}$ThO ~\cite{Liang1995,Hammond2002,Flambaum2019,Chen2024,Ng2025}, and nuclear hyperfine structure (HFS) measurements to benchmark the electronic orbital calculations on the EDM-sensitive \Hh ~state in ThO~\cite{Fleig2014,Skripnikov2015}. Improved HFS calculations on ThO are now underway using the relativistic coupled-cluster method~\cite{Haase2020}.  This work thus provides both the first demonstration of optical cycling on a radioactive molecule and several impactful immediate applications. It also motivates further investigation on optical-cycling of other non-traditional species. We note that this is the first optical cycling that has been quantified on a neutral molecule not of alkali-atom-like or alkaline-earth-atom-like structure.

The optical cycle demonstrated here is limited by the loss to the metastable electronic states \Hh, $v=0$ and \Q, $v=0$~\cite{Kokkin2014_ULI}. Future work will examine how to integrate these states into the optical cycle. These states differ in electronic character from the ground state and are highly sensitive to stray fields. Repumping these states will require departure from the typical optical cycling paradigm, and will continue to elucidate how the cycling process can be extended to molecules of more general structures. If these states, as well as $X,v=1$ can be successfully repumped, the optical cycle will be sufficiently closed for one- or two-dimensional transverse cooling. Such an advancement may potentially lead to further improved beam experiments using ThO for tests of fundamental physics, and would help elucidate the path to quantum control of a much wider range of molecular species than is currently accessible. 

\section*{Acknowledgments}
The authors thank David DeMille, Gerald Gabrielse, and John M. Doyle for insightful discussions, and for their generous support of allowing us to use the legacy ACME thorium beamline. The authors also thank Zack Lasner, Grace K. Li, Olivier Grasdijk, Ayami Hiramoto, Takahiko Masuda, Collin Diver, Peiran Hu, Zhen Han, Daniel G. Ang, Jaideep T. Singh, and Patrick R Stollenwerk for helpful conversations.

\section*{Funding}
This material is based upon work supported by the U.S. Department of Energy, Office of Science, Office of Nuclear Physics and used resources of the Facility for Rare Isotope Beams (FRIB) Operations, which is a DOE Office of Science User Facility under Award Number DE-SC0023633.

\section*{Author Contributions}
A.F., D.G., N.E. and X.W. designed the experimental methodology and conducted investigation. Simulations designed by D.G. and X.W. All authors discussed and validated results. A.F. and X.W. wrote, and all authors edited and reviewed, the manuscript. Project is managed by X.W.

\section*{Data Availability}
Both experimental and simulation data are available upon reasonable request.

\printbibliography

\pagebreak
\appendix
\section{Rate Equation Model\label{App1}}
In this appendix, we first describe in detail the construction of the rate-equation model discussed in the main text (Sec.~\ref{subsec:Rsc}). Then, we describe what this model informs us about the cycling process in different experimental regimes, i.e. the different relative ordering of the rates, $\Gamma, R_{ps},\,\mathrm{and}~\Omega$. As noted in the main text, we draw from Refs.~\cite{Wall2008,Hofsss2021,FITCH2021157} to set up the model. 

\subsection{Making the Model}
We first define our basis as the set of all magnetic sublevels in each relevant rotational level. The population of each state is given a separate variable. For a ground state $i\in \{M_J''\}$, we denote the fractional population $g_i$; analogously $e_j$ is the population in an excited state $j\in \{M_J'\}$. The essence of a rate-equation model (as opposed to, say, OBEs) is that coherences are completely ignored: the population only moves between states in the basis as a result of a) stimulated optical pumping or b) spontaneous emission. Each of these terms is used to construct first-order differential equations of each state's population with respect to time.

Optical pumping is introduced as rates $R_{ij} = |\Omega_{ij}|^2/\Gamma$. The Rabi frequency is defined in the typical way, $\Omega_{ij}=d_{ij}E/\hbar$, where $d_{ij}$ is the polarization-dependent transition dipole moment (TDM) and $E = \sqrt{\frac{2I}{c\epsilon_0}}$ for light of intensity $I$. These TDMs are polarization dependent, so by the Wigner-Eckhart theorem we can rewrite $d_{ij}=S_{ij}D$, where $D$ is the reduced (projection-independent) TDM (1.84 D) and $S_{ij}$ is a rotational factor given in Eq.~\ref{eq:Sfactor} in text. 

The form of these pumping rates is derived from perturbation theory in the case of a weak pump field, and so technically is only correct when $\Omega_{ij}\ll\Gamma$. Nevertheless, in the case of optical cycling with a single laser beam (no overlapping retroreflection), coherences do not have significant physical effect and essentially average to the rate equation result, even at much larger $\Omega$.

Spontaneous emission (SE) rates are included in a similar manner. Since the SE out of an excited state is irrespective of final state of the decay, the SE out of an excited state $j$ occurs at $-\Gamma$. The rate into a ground state $i$ from a specific excited state depends on the rate out of the excited state and the total branching fraction. It can also be written in terms of $S$ factors as $b\Gamma S_{ij}^2/\sum_kS_{kj}^2)$, where $b$ is the vibronic branching fraction and the latter sum is over all ground states to ensure normalization.

To clarify the preceding paragraphs, we present the equations for a $Q(1)$ model with $\sigma^+$ polarized light. We decompose $\Omega_{ij} = S_{ij}\Omega$ and use numerical values for $S_{ij}$, but otherwise leave the equations in symbolic form.

\begin{gather*}
    \frac{dg_{-1}}{dt} = \frac{|\Omega|^2}{4\Gamma}(e_{0} - g_{-1})+\frac{b\Gamma}{2}(e_0 + e_{-1})\\
    \frac{dg_{0}}{dt} = \frac{|\Omega|^2}{4\Gamma}(e_{+1} - g_{0})+\frac{b\Gamma}{2}(e_0 + e_{+1})\\
    \frac{dg_{+1}}{dt} = \frac{b\Gamma}{2}(e_1 + e_{0})
    \\
    \frac{de_{-1}}{dt} = -\frac{\Gamma}{2}(e_{-1});\\
    \frac{de_{0}}{dt} = \frac{|\Omega|^2}{4\Gamma}(g_{-1} - e_{0})-\frac{\Gamma}{2}(e_0 + e_{+1});\\
    \frac{de_{+1}}{dt} = \frac{|\Omega|^2}{4\Gamma}(g_{0} - e_{+1}) -\frac{\Gamma}{2}(e_{+1});
\end{gather*}

The dark state, $g_{+1}$, is obvious. By symmetry, it is clear that swapping to $\sigma^-$ polarization will change dark states to $g_{-1}$. To implement polarization switching, we simply swap these equations with their symmetric $\sigma^-$ form at the desired rate during the numerical solution. We also construct a $Q(2)$ model, but do not present the equations explicitly here, since the concepts are the same.

The instantaneous scattering rate, recall, is equivalent to the spontaneous emission rate. At the ensemble level, this is strictly given by $\Gamma\times(\sum_j e_j)$, since this is the number of photons emitted per unit time. However, typically these models are meant to elucidate the dynamics of an average particle in the cycle, which means we should not count the population that has been lost. Thus the scattering rate is better described as $\Gamma\times(\sum_j e_j)/[(\sum_j e_j)+(\sum_{i} g_i)]$, which gives the rate at which the cycling molecules continue to scatter photons, on average. Clearly, if the loss out of the cycle is low, these equations are nearly equivalent.

\subsection{Using the model}

We use the rate equation model to qualitatively assess the effect of different relative orderings of $\Gamma, \Omega$, and $R_{ps}$~(See Tab.~\ref{tab:rate_eq_mod}). In general, polarization switching is expected to constrain $R_{sc}$ the most severely when $R_{ps}<\Gamma\lesssim\Omega$; when molecules have time to be optically pumped into the dark states but not spend excessive time there before the polarization changes\footnote{The rate equation model may not be exactly quantitative in this intensity regime. However, since there are no coherent effects in an ordinary cycling context, transient coherences should average out. This conclusion is supported by similar work in other molecules~\cite{Wall2008,Hofsss2021}}. This expectation is confirmed by the model. At $I=10I_{sat}$, we find the average $R_{sc}$ over 5 us (chosen to be longer than initial transient effects) increases over $6$ times when $R_{ps}$ increases from $0$ to $1$ \Msi; from $0.26\times10^6$ s$^{-1}$ to $1.7\times10^6$ s$^{-1}$. From the model, it is clear that this difference is almost entirely due to less time spent in dark states. As such, $R_{sc}$ approaches a limiting value at high intensity: when $I\approx100I_{sat}$ (with $R_{ps}=1$~\Msi), $R_{sc}$ only reaches slightly above $2.2\times10^6$ s$^{-1}$, despite an over three times larger $\Omega$ than when $I=10I_{sat}$. 

Closer to $\Gamma=\Omega$, the switching matters less, owing to significant population being still ``bright'' at the switching time. The model again confirms this: when $I=I_{sat}$, increasing $R_{ps}$ from $0$ to $1$~\Msi~only increases $R_{sc}$ from $0.23\times10^6$ s$^{-1}$ to $0.48\times10^6$ s$^{-1}$. 

In all of the above cases, the ultimate scattering rates only increase slightly when $R_{ps}$ increases from $0.5$ to $3$~\Msi. Below $R_{ps}=0.5$~\Msi, $R_{sc}$ tends to decrease as population spends a larger fraction of the time in dark states even at low pumping rates. Above 3~\Msi, $R_{ps}$ becomes comparable to $\Gamma/2$. We do not experimentally probe $R_{ps}$ this large, but expect it not to have a significantly different effect as long as both $R_{ps},\Gamma<\Omega$. More investigation of this regime is warranted in the future, since several unorthodox species proposed for optical cycling have very long excited state lifetimes (e.g. CH\cite{Schnaubelt2021}, or AuC~\cite{Augenbraun2020}), and can trivially reach $R_{ps}\approx\Gamma$.

For completeness, if $\Omega\ll\Gamma$, $R_{sc}$ is clearly limited by excitation time, regardless of $R_{ps}$. We do not study this case in depth, since most applications of cycling require fast scattering rates.

\begin{table}[!h]
    \centering
    \caption{Comparison between the scattering rates $R_{sc}$ of $Q(1)$ and $Q(2)$ transitions, modeled by the rate equation as a function of polarization switching rate $R_{ps}$ and laser intensity $I$($\propto\Omega^2$). Note that both parameters are varied on a logarithmic scale. Spontaneous emission rate is fixed to $\Gamma = 1/(115$~ns) here. At low intensity, $R_{ps}$ makes little difference, but as intensity increases, faster $R_{ps}$ are necessary to maximize $R_{sc}$.\\}
    \begin{tabular}{|c|c|c|c|c|c|c|c|c|}
        \hline
        $[R_{sc}(Q(1))~|~R_{sc}(Q(2))]~(10^6~\textrm{s}^{-1})$  & \multicolumn{2}{c|}{$I=0.1I_{sat}$} & 
         \multicolumn{2}{c|}{$I=I_{sat}$} & \multicolumn{2}{c|}{$I=10I_{sat}$} &  \multicolumn{2}{c|}{$I=100I_{sat}$}  \\
        \hline
        $R_{ps}=0.1$~MHz & $0.058$ & $0.073$ & $0.23$&$0.37$ & $0.26$&$0.45$ & $0.26$&$0.45$ \\
        \hline
        $R_{ps}=0.3$~MHz & $0.063$&$0.076$ & $0.44$&$0.60$ & $1.0$&$1.8$ & $1.0$&$2.0$ \\
        \hline
        $R_{ps}=1$~MHz & $0.063$&$0.077$ & $0.48$&$0.63$ & $1.7$&$2.4$ & $2.2$&$3.3$  \\
        \hline
        $R_{ps}= 3 $~MHz &  $0.063$&$0.077$ & $0.48$&$0.63$ & $1.9$&$2.5$ & $2.6$&$3.5$\\
        \hline
        $R_{ps} = 10$~MHz & $0.063$&$0.077$ & $0.48$&$0.64$ & $1.9$&$2.5$ & $2.8$&$3.6$ \\
        \hline
    \end{tabular}
    \label{tab:rate_eq_mod}
\end{table}

\end{document}